\documentclass[prl,twocolumn,floatfix,groupedaddress,nofootinbib,showpacs,preprintnumbers,
amsmath,amssymb,amsfonts,superscriptaddress,widetable] {revtex4}
\usepackage{graphicx}% Include figure files
\usepackage{dcolumn}%Align table columns on decimal point
\usepackage{mathrsfs}
\usepackage{bm}% bold math

\usepackage{graphicx}% Include figure files
\usepackage{dcolumn}% Align table columns on decimal point
\usepackage{bm}% bold math
\usepackage[usenames]{color}
\begin{document}

\title{Relativistic models of the neutron-star matter equation of state}

\author{F.~J.~Fattoyev}\email{ff07@fsu.edu}
\author{J.~Piekarewicz}~\email{jpiekarewicz@fsu.edu}
\affiliation{Department of Physics, Florida State University, 
                  Tallahassee, Florida 32306, USA}

\date{\today}

\begin{abstract}
 Motivated by a recent astrophysical measurement of the pressure 
 of cold matter above nuclear-matter saturation
 density~\cite{Ozel:2010fw}, we compute the equation of state of
 neutron star matter using accurately calibrated relativistic
 models. The uniform stellar core is assumed to consist of
 nucleons and leptons in beta equilibrium; no exotic degrees of 
 freedom are included. We found the predictions of these models to be
 in fairly good agreement with the measured equation of state.  Yet the
 {\sl Mass-vs-Radius} relations predicted by these same models display
 radii that are consistently larger than the observations. 
 %At present, we offer no solution to this conflict.
\end{abstract}
\pacs{26.60.-c,26.60.Kp,21.60.Jz}
\maketitle

The quest for the Holy Grail of Nuclear Physics---the equation of
state (EOS) of hadronic matter---remains an area of intense activity
that cuts across a variety of disciplines. Indeed, the limits of
nuclear existence, the dynamics of heavy-ion collisions, the structure
of neutron stars, and the collapse of massive stellar cores all depend
sensitively on the equation of state.  With the advent and
commissioning of sophisticated new radioactive beam facilities,
powerful heavy-ion colliders, telescopes operating at a variety of
wavelengths, and more sensitive gravitational wave detectors, one will
be able to probe the nuclear dynamics over a wide range of nucleon
asymmetries, temperatures, and densities.  However, in the present
contribution we focus on the dynamics of cold matter under extreme
conditions of density (both small and large) and for this case neutron
stars remain the tool of
choice~\cite{Lattimer:2000nx,Lattimer:2004pg,Lattimer:2006xb}.  Being
both very compact and extremely dense, neutron stars are unique
laboratories for probing the equation of state of neutron-rich matter
under conditions unattainable by terrestrial experiments.

Intimately connected to the equation of state of cold, neutron-rich
matter is the {\sl mass-vs-radius} ({\sl M-R}) relationship of neutron
stars. Indeed, an EOS is the sole ingredient that must be supplied to
solve the equations of stellar structure ({\sl i.e.,} the
Tolman-Oppenheimer-Volkoff equations). Conversely, knowledge of the {\sl M-R}
relation is sufficient to uniquely determine the equation of state of
neutron star matter~\cite{Lindblom:1992}.  As argued by Lindblom
almost 20 years ago, the availability of such information---even from
a single neutron star---will provide interesting information about
the equation of state~\cite{Lindblom:1992}.  Viewed in this light, the
recent report of combined mass-radius measurement for three neutron
stars and the subsequent determination of the equation of state is
significant~\cite{Ozel:2010fw}. In particular, the conclusion that the
EOS so determined is softer than those containing only nucleonic 
degrees of freedom is both interesting and provocative.

In this contribution we compute the equation of state of neutron star
matter and the resulting {\sl M-R} relation using
accurately-calibrated relativistic mean-field models. These models
have been calibrated to the properties of infinite nuclear matter at
saturation density~\cite{Mueller:1996pm}, to the ground-state
properties of finite nuclei~\cite{Lalazissis:1996rd,Lalazissis:1999},
or to both~\cite{Todd-Rutel:2005fa}.  Unlike the former two, the
latter parametrization predicts a significantly soft symmetry energy,
a feature that appears consistent with the behavior of dilute neutron
matter (see Ref.~\cite{Piekarewicz:2009gb} and references therein).  A
detailed explanation of the role of the model parameters on the
equation of state is given below. We note, however, that none of the
models considered in this work include exotic degrees of freedom, such
as hyperons, meson (condensates), or quarks. In this regard, our
results are mixed when compared with the conclusions of
Ref.~\cite{Ozel:2010fw}. On the one hand, the stellar radii predicted
by the relativistic models are larger than observed, seemingly
confirming that such equations of state are too stiff. On the other
hand, the agreement between the predicted and observed EOS
suggests the opposite.

The structure of neutron stars is sensitive to the equation of state
of cold, fully catalyzed, neutron-rich matter over an enormous range
of densities~\cite{Lattimer:2000nx,Lattimer:2004pg,Lattimer:2006xb}.
For the low-density outer crust we employ the equation of state of
Baym, Pethick, and Sutherland~\cite{Baym:1971pw}.  At densities of
about a third to a half of nuclear-matter saturation density,
uniformity in the system is restored and for this (liquid-core) region
we use an EOS derived from a representative set of accurately
calibrated relativistic mean-field
models~\cite{Mueller:1996pm,Lalazissis:1996rd,Lalazissis:1999,
Todd-Rutel:2005fa}. It has been speculated that the region between the
outer crust and the liquid core consists of complex and exotic
structures, collectively known as {\sl{nuclear
pasta}}~\cite{Ravenhall:1983uh,Hashimoto:1984,Lorenz:1992zz}.  Whereas
significant progress has been made in simulating this exotic
region~\cite{Horowitz:2004yf,Horowitz:2004pv,Horowitz:2005zb}, a
detailed equation of state is still missing. Hence, we resort to a
fairly accurate polytropic EOS to interpolate between the solid crust
and the uniform liquid interior~\cite{Link:1999ca,Carriere:2002bx}.
To compute the transition density from the liquid core to the solid
crust we employ a relativistic random-phase-approximation (RPA)
analysis to search for the critical density at which the uniform
system becomes unstable to small amplitude density
oscillations~\cite{Carriere:2002bx}.

Accounting for most of the stellar radius and practically all of its
mass, the liquid core is structurally the most important component of
the star.  Matter in the liquid core is assumed to be composed of
neutrons, protons, electrons, and muons in chemical equilibrium.  We
reiterate that no exotic degrees of freedom are included in the model.
Both electrons and muons are treated as non-interacting relativistic
Fermi gases. For the hadronic component, the equation of state is
generated using accurately-calibrated relativistic models. Details on
the calibration procedure may be found in
Refs.~\cite{Serot:1984ey,Serot:1997xg,Horowitz:2000xj, Todd:2003xs}.
The model includes a nucleon field ($\psi$) interacting via
standard Yukawa couplings to two isoscalar mesons (a scalar $\phi$ and
a vector $V^{\mu}$) and one vector-isovector meson
($b^{\mu}$)~\cite{Serot:1984ey,Serot:1997xg}.  Such an interacting
Lagrangian density may be written as 
follows~\cite{Serot:1984ey, Serot:1997xg,Mueller:1996pm}:
\begin{equation}
{\mathscr L}_{\rm int} =
\bar\psi \left[g_{\rm s}\phi   \!-\!
         \left(g_{\rm v}V_\mu  \!+\!
    \frac{g_{\rho}}{2}\mbox{\boldmath$\tau$}\cdot{\bf b}_{\mu}
          \right)\gamma^{\mu} \right]\psi -
          U(\phi,V^{\mu},{\bf b^{\mu}}) \;.
\label{Lagrangian}
\end{equation}
In addition to the Yukawa couplings ($g_{\rm s}$, $g_{\rm v}$, and
$g_{\rho}$), the model is supplemented by non-linear meson 
interactions given by
%%%
\begin{eqnarray}
  U(\phi,V^{\mu},{\bf b}^{\mu}) &=&
   \frac{\kappa}{3!} (g_{\rm s}\phi)^3 \!+\!
    \frac{\lambda}{4!}(g_{\rm s}\phi)^4  \!-\!
   \frac{\zeta}{4!}
    \Big(g_{\rm v}^2 V_{\mu}V^\mu\Big)^2 
        \nonumber \\ & \!-\! &
    \Lambda_{\rm v}
    \Big(g_{\rho}^{2}\,{\bf b}_{\mu}\cdot{\bf b}^{\mu}\Big)
    \Big(g_{\rm v}^2V_{\mu}V^\mu\Big) \;.
\label{USelf}
\end{eqnarray}
%%%
The inclusion of scalar cubic ($\kappa$) and quartic ($\lambda$)
self-interactions dates back to the late seventies~\cite{Boguta:1977xi}
and is instrumental for softening the incompressibility coefficient of
symmetric nuclear matter, as required to explain the excitation of the
the nuclear {\sl breathing mode}~\cite{Youngblood:1999}.

Of particular interest and of critical importance to the present study
are the vector self-interaction ($\zeta$) and the {\sl
isoscalar-isovector} mixing term $\Lambda_{\rm
v}$~\cite{Horowitz:2000xj,Todd-Rutel:2005fa}.  That both of these
parameters are zero in the enormously successful NL3 model (see
Table~\ref{Table1}) suggests that existing laboratory data are fairly
insensitive to the physics encoded in these two parameters.  Indeed,
M\"uller and Serot found possible to build models with different
values of $\zeta$ that reproduce the same observed properties at
normal nuclear densities, yet produced maximum neutron star masses
that differ by almost one solar mass~\cite{Mueller:1996pm}.  This
result indicates that observational data on neutron stars---rather
than laboratory experiments---may provide the only meaningful
constraint on the high-density component of the equation of
state. Further, it indicates that the empirical parameter $\zeta$
provides an efficient tool to control the high-density component of
the equation of state.  

The isoscalar-isovector coupling constant $\Lambda_{\rm v}$ was
introduced in Ref.~\cite{Horowitz:2000xj} to modify the poorly known
density dependence of the symmetry energy. The symmetry energy
represents the energy cost involved in changing protons into neutrons
(and vice-versa). To a good approximation, it is given by the
difference in energy between pure neutron matter and symmetric nuclear
matter.  With only one isovector parameter ($g_{\rho}$) to adjust,
relativistic mean-field models have traditionally predicted a {\sl
stiff} symmetry energy. The addition of $\Lambda_{\rm v}$ provides a
simple---yet efficient and reliable---method of softening the symmetry
energy without compromising the success of the model in reproducing
well determined ground-state observables~\cite{Todd-Rutel:2005fa}.
Indeed, whereas models with different values of $\Lambda_{\rm v}$
reproduce the same exact properties of symmetric nuclear matter, they
yield vastly different predictions for both the neutron radii of heavy
nuclei and for the radius of neutron
stars~\cite{Horowitz:2000xj,Horowitz:2001ya}.  Given that the
neutron star radius is believed to be primarily controlled by the
symmetry pressure at intermediate densities~\cite{Lattimer:2006xb},
the upcoming Parity Radius Experiment (PREx) at the Jefferson
Laboratory ({\sl with an imminent start date of March, 2010}) 
will provide a unique laboratory constraint on a fundamental 
neutron star property~\cite{Horowitz:1999fk,Michaels:2005}. 

In summary, the two empirical parameters $\zeta$ and $\Lambda_{\rm v}$
provide a highly economical and efficient control of the softness of
the high-density component of equation of state and of the symmetry
pressure at intermediate densities, respectively---with the former
primarily controlling the maximum neutron star mass and the latter the
stellar radius. Parameter sets for all the models employed in this
work are listed in Table~\ref{Table1}.

%%%%%%%%%%%%%%%%%%%%%%%%%%%%%%%%%%%%%%%%%%%%%%%%%%%%%%%%%%%%%%%%%
\begin{widetext}
\begin{center}
\begin{table}[t]
\begin{tabular}{|l||c|c|c|c|c|c|c|c|c|c|}
 \hline
 Model & $m_{\rm s}$  & $m_{\rm v}$  & $m_{\rho}$  
       & $g_{\rm s}^2$ & $g_{\rm v}^2$ & $g_{\rho}^2$
       & $\kappa$ & $\lambda$ & $\zeta$ & $\Lambda_{\rm v}$\\
 \hline
 \hline
 NL3  & 508.194 & 782.501 & 763.000 & 104.3871 & 165.5854 &  79.6000 
         & 3.8599  & $-$0.01591 & 0.00 & 0.00 \\
 MS    & 485.000 & 782.500 & 763.000 & 111.0426 & 216.8998 &  70.5941 
          & 0.5082  & $+$0.02772 & 0.06 & 0.00 \\
 FSU   & 491.500 & 782.500 & 763.000 & 112.1996 & 204.5469 & 138.4701 
          & 1.4203  & $+$0.02376 & 0.06 & 0.03 \\
  XS    & 491.500 & 782.500 & 763.000 & 131.0059 & 258.1044 & 213.9596 
          & 0.0079  & $+$0.04339 & 0.09 & 0.04 \\
\hline
\end{tabular}
\caption{Parameter sets for the four models used in the text to generate
  the equation of state. The parameter $\kappa$ and the meson 
  masses $m_{\rm s}$, $m_{\rm v}$, and $m_{\rho}$ are all given in MeV. The
nucleon mass has been fixed at $M\!=\!939$~MeV in all the models.}
\label{Table1}
\end{table}
\end{center}
\end{widetext}
%%%%%%%%%%%%%%%%%%%%%%%%%%%%%%%%%%%%%%%%%%%%%%%%%%%%%%%%%%%%%%%%%

\begin{figure}[tb]
\vspace{-0.05in}
\includegraphics[width=1.00\columnwidth,angle=0]{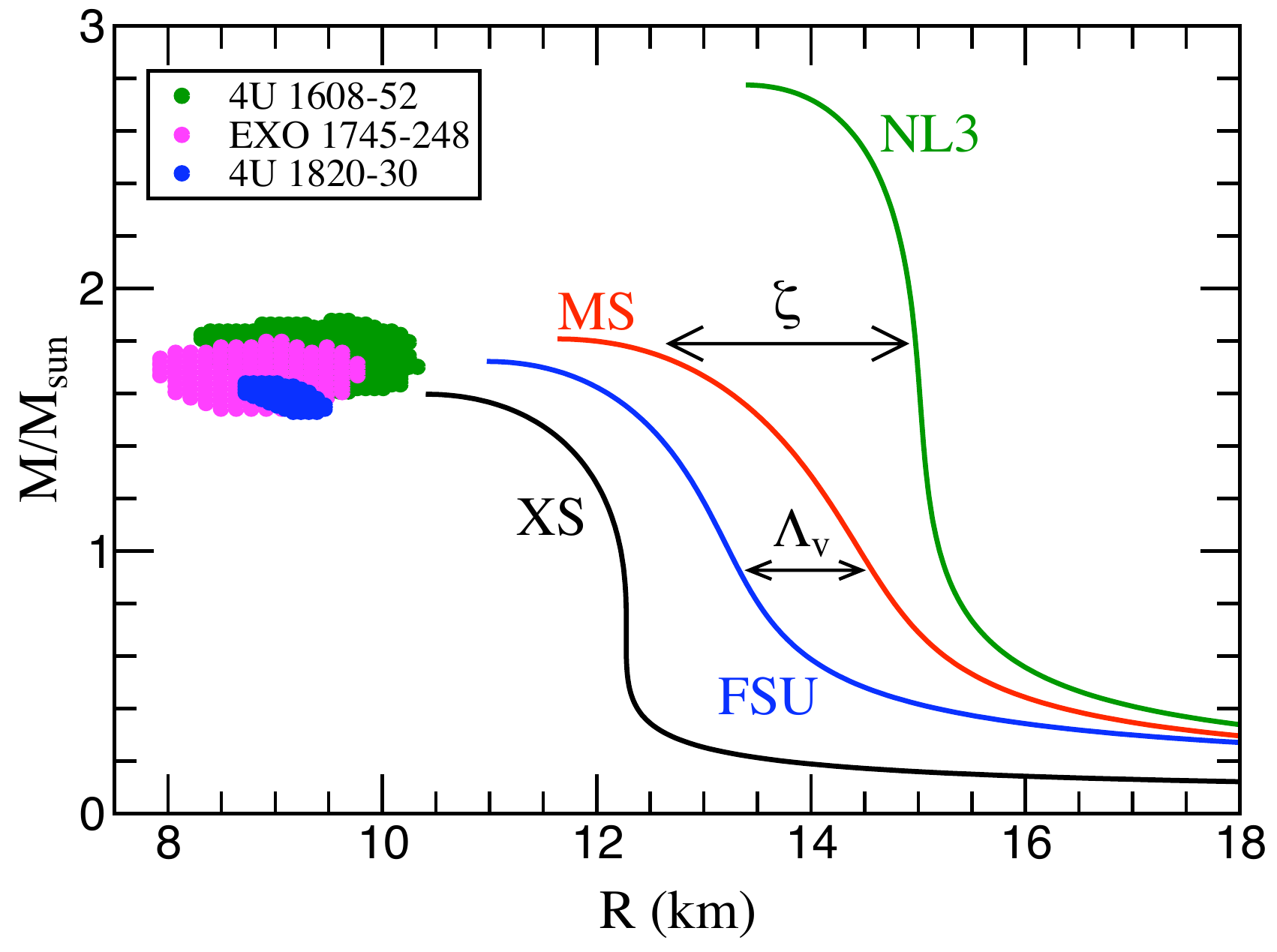}
\caption{(Color online) {\sl Mass-vs-Radius} relation predicted by 
 the four relativistic mean-field models discussed in the text. The
 observational data represent 1$\sigma$ confidence contours for
 the three neutron stars reported in Ref.~\cite{Ozel:2010fw}}.
\label{Fig1}
\end{figure}

In Fig.~\ref{Fig1} we compare observational results for three neutron
star masses and radii against the model predictions. These neutron
stars are in the binaries 4U~1608-52~\cite{Guver:2008gc},
EXO~1745-248~\cite{Ozel:2008kb}, and
4U~1820-30~\cite{Guver:2010td}. The very stiff behavior of the NL3
equation of state is immediately evident.  With both empirical
parameters $\zeta$ and $\Lambda_{\rm v}$ set equal to zero, it is not
surprising that the NL3 model predicts neutron star masses as large as
$2.8~M_{\odot}$ with very large radii. As compared to the
observational data, the NL3 model suggests a radius for a 1.7
solar-mass neutron star that is about 6~km too large.  Moreover, the
NL3 equation of state is so stiff that gravity in a $2.8~M_{\odot}$
neutron star can compress matter to only about four times normal
nuclear density (see Table~\ref{Table2}). All these, even when the
model provides an excellent description of many laboratory
observables.

%%%%%%%%%%%%%%%%%%%%%%%%%%%%%%%%%%%%%%%%%%%%%%%%%%%%%%%%%%%%%%%%%
%%%%%%%%%%%%%%%%%%%%%%%%%%%%%%%%%%%%%%%%%%%%%%%%%%%%%%%%%%%%%%%%%

\begin{figure}[tb]
\vspace{-0.05in}
\includegraphics[width=1.00\columnwidth,angle=0]{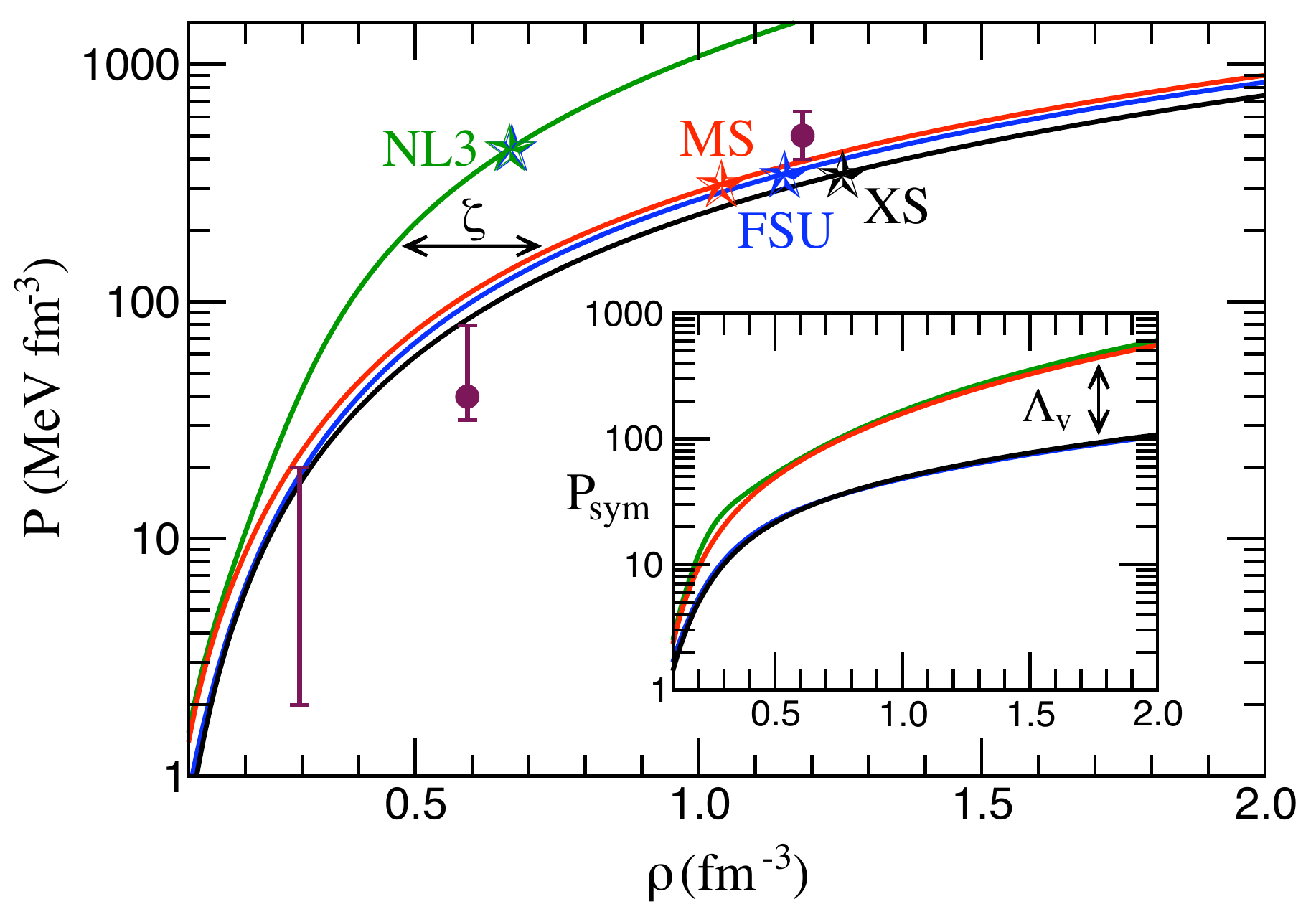}
\caption{(Color online) Equation of state ({\sl Pressure vs
 baryon density)} of neutron star matter predicted by the 
 four relativistic mean-field models discussed in the text.
 The three data point are from the observational 
 extraction as described in Ref.~\cite{Ozel:2010fw}. 
 The symbols (stars) indicate the central density and pressure for
 the maximum-mass neutron star. The inset shows the symmetry 
 pressure, given as the pressure  of pure neutron matter minus 
 that of symmetric nuclear matter.}
\label{Fig2}
\end{figure}

As first suggested by M\"uller and Serot~\cite{Mueller:1996pm}, adding
a vector self-interaction (with $\zeta\!=\!0.06$) dramatically reduces
the repulsion at high densities and ultimately the limiting
neutron star mass. As compared to the NL3 parameter set, the maximum
neutron star mass predicted by M\"uller and Serot (MS) is reduced by
almost one solar mass (see Fig.~\ref{Fig1} and Table~\ref{Table2}).
Consistent with this softening is a significant increase in the
compactness of the star.  For example, for a neutron star mass of
$1.8~M_{\odot}$, NL3 predicts a stellar radius that is more than 3
kilometers larger than MS.  Note, however, that the density dependence
of the symmetry energy predicted by NL3 and MS is practically
identical (see inset in Fig.~\ref{Fig2}). In particular, this is
reflected in the identical prediction of 0.28~fm for the neutron-skin
thickness of ${}^{208}$Pb. This suggests that tuning the density
dependence of the symmetry energy---via the addition of the
isoscalar-isovector mixing term $\Lambda_{\rm v}$---may yield a
further reduction in neutron star radii~\cite{Horowitz:2001ya}, as
suggested by observation.

%%%%%%%%%%%%%%%%%%%%%%%%%%%%%%%%%%%%%%%%%%%%%%%%%%%%%%%%%%%%%%%%%
  \begin{table}
  \begin{tabular}{|c|c|c|c|c|c|}
    \hline
 Model & $\rho$ & $P$ & $M$ & $R$ & $R_{1.4}$ \\
 \hline
 \hline
 NL3   & 0.667 & 440.58 & 2.78 & 13.39 & 15.05 \\
 MS     & 1.040 & 311.92 & 1.81 & 11.64 & 13.78 \\
 FSU    & 1.153 & 345.78 & 1.72 & 10.97 & 12.66 \\
 XS      & 1.252 & 345.37 & 1.60 & 10.41 & 11.73 \\
 \hline
 \end{tabular}
 \caption{Predictions for the central baryon density, 
   central pressure, mass, and radius of the limiting 
   neutron star for the four models employed in the text.
   The last column lists predictions for the radius of 
   a 1.4 solar-mass neutron star. The baryon
   density is given in ${\rm fm}^{-3}$, the
   pressure in ${\rm MeV}~{\rm fm}^{-3}$, the mass 
   in solar masses, and the radii in kilometers.}
\label{Table2}
 \end{table}
%%%%%%%%%%%%%%%%%%%%%%%%%%%%%%%%%%%%%%%%%%%%%%%%%%%%%%%%%%%%%%%%%

Incorporating information on nuclear collective modes in the
calibration procedure of the FSUGold model favors a non-zero value
for $\Lambda_{\rm v}$~\cite{Todd-Rutel:2005fa}.  Further, it now seems
that the resulting softening of the symmetry energy is consistent with
the EOS of dilute neutron matter predicted by various microscopic
approaches (see Refs.~\cite{Schwenk:2005ka,Gezerlis:2009iw,
Piekarewicz:2007dx, Piekarewicz:2009gb} and references therein). That
the addition of $\Lambda_{\rm v}$ produces the intended effect can be
appreciated in Fig.~\ref{Fig1} and Table~\ref{Table2}. That is,
although one has adopted the same value of $\zeta$ for both MS and
FSUGold, their predictions for the radius of a {\sl ``canonical''} 1.4
solar-mass neutron star differ by more than one kilometer. Related to
this fact is the significantly smaller neutron-skin thickness of
${}^{208}$Pb predicted by FSUGold (0.21~fm {\sl vs} 0.28~fm).
However, it appears that the combined softening of the EOS at
high densities (through $\zeta$) and of the symmetry pressure (through
$\Lambda_{\rm v}$) is insufficient to explain the observational data;
the minimum stellar radius predicted by the FSUGold model is about 11
km, significantly larger than suggested by observation.

In an effort to describe the observational data, we have constructed
an {\sl ``Extra Soft''}~(XS) relativistic mean-field model constrained
by the properties of symmetric nuclear matter at saturation density
({\sl i.e.,} equilibrium density, binding energy per nucleon, and
incompressibility coefficient). In regards to these properties, the
model is indistinguishable from FSUGold.  The only additional
constraint imposed on the model is that its limiting mass be no
smaller than 1.6 solar masses. We feel that lowering this limiting
value any further may start conflicting with the 
observational data.  Although no exhaustive
parameter search was conducted, we trust that the resulting {\sl
extra-soft} equation of state (as given in Table~\ref{Table1}) is
representative of the softness that may be achieved with present-day
relativistic mean-field models. With such a soft model, neutron star
radii get significantly reduced indeed (see Fig.~\ref{Fig1}). For
example, the neutron radius of a 1.4~$M_{\odot}$ neutron star is
reduced by almost one kilometer relative to the FSUGold prediction
(see Table~\ref{Table2}) and by more than 1.5~km at its limiting mass
of 1.6~$M_{\odot}$.  Still, the minimum neutron star radius of
$R\!=\!10.41$~km predicted by the model remains outside the 
reported $1\sigma$ confidence contours~\cite{Ozel:2010fw}.

Do we then conclude that the results presented in Fig.~\ref{Fig1} are
indicative of relativistic equations of state that are too stiff? Do
the observational results unambiguously called for a softer equation
of state, as would be produced by exotic states of matter, such as
meson condensates and/or quark matter?  To answer this question we
compare in Fig.~\ref{Fig2} the various equations of state used to
generate Fig.~\ref{Fig1} against the values extracted from the
observational data~\cite{Ozel:2010fw}. The inset in the figure
displays the symmetry pressure for the models under consideration. The
observed softening of the symmetry pressure between models is entirely
due $\Lambda_{\rm v}$.  Note, however, that unlike the neutron-skin
thickness of neutron-rich nuclei, the radius of a neutron star is
not uniquely constrained by the symmetry pressure at low to
intermediate densities~\cite{Horowitz:2001ya}. Thus models with
similar symmetry pressures may---and do---predict significantly
different stellar radii.  Contrary to the expectations generated by
Fig.~\ref{Fig1}, most of the equations of state {\sl are not too
stiff}.  Indeed, with the exception of NL3, the remaining equations of
state appear, if anything, slightly {\sl too soft} at the highest
density. Based on these results---and these results alone---nucleonic
equations of state do not seem to be in conflict with the
observational data.

%%%%%%%%%%%%%%%%%%%%%%%%%%%%%%%%%%%%%%%%%%%%%%%%%%%%%%%%%%%%%%%%%

%%%%%%%%%%%%%%%%%%%%%%%%%%%%%%%%%%%%%%%%%%%%%%%%%%%%%%%%%%%%%%%%%

In summary, the {\sl Mass-vs-Radius} relation of neutron stars was
computed using equations of state derived from relativistic mean-field
models.  Although the models are calibrated in the vicinity of
nuclear-matter saturation density, it is possible to tune their
high-density behavior in a highly efficient and economical manner.  In
this contribution we have used two parameters to control the maximum
neutron star mass and the stellar radius.  As we compared our
predictions to the observational data a conflict emerged. Whereas one
could generate equations of state that are in agreement with
observation, the predicted stellar radii are too large.  This result
is particularly intriguing given that {\sl ``inversion''} methods
exist for extracting the equation of state of stellar matter directly
from masses and radii of neutron stars~\cite{Lindblom:1992}.  Thus,
one would expect that if the {\sl M-R} predictions do not match
observation, neither would the equations of state.  Clearly, to
reconcile these facts much work remains to be done in both the
observational and theoretical fronts. For now we must
conclude---although the existence of exotic stars is very
appealing---that the downfall of the purely nucleonic equations of
state may be premature.

%%%%%%%%%%%%%%%%%%%%%%%%%%%%%%%%%%%%%%%%%%%%%%%%%%%%%%%%%%%%%%%%%
\begin{acknowledgments}
 The authors want to thank Prof. F. \"Ozel for making available 
 the observational data. This work was supported in part by a 
 grant from the U.S. Department of Energy DE-FD05-92ER40750. 
\end{acknowledgments}

\bibliography{MassRadius.bbl}

\begin{thebibliography}{34}
\expandafter\ifx\csname natexlab\endcsname\relax\def\natexlab#1{#1}\fi
\expandafter\ifx\csname bibnamefont\endcsname\relax
  \def\bibnamefont#1{#1}\fi
\expandafter\ifx\csname bibfnamefont\endcsname\relax
  \def\bibfnamefont#1{#1}\fi
\expandafter\ifx\csname citenamefont\endcsname\relax
  \def\citenamefont#1{#1}\fi
\expandafter\ifx\csname url\endcsname\relax
  \def\url#1{\texttt{#1}}\fi
\expandafter\ifx\csname urlprefix\endcsname\relax\def\urlprefix{URL }\fi
\providecommand{\bibinfo}[2]{#2}
\providecommand{\eprint}[2][]{\url{#2}}

\bibitem[{\citenamefont{Ozel et~al.}(2010)\citenamefont{Ozel, Baym, and
  Guver}}]{Ozel:2010fw}
\bibinfo{author}{\bibfnamefont{F.}~\bibnamefont{Ozel}},
  \bibinfo{author}{\bibfnamefont{G.}~\bibnamefont{Baym}}, \bibnamefont{and}
  \bibinfo{author}{\bibfnamefont{T.}~\bibnamefont{Guver}}
  (\bibinfo{year}{2010}), \eprint{1002.3153}.

\bibitem[{\citenamefont{Lattimer and Prakash}(2001)}]{Lattimer:2000nx}
\bibinfo{author}{\bibfnamefont{J.~M.} \bibnamefont{Lattimer}} \bibnamefont{and}
  \bibinfo{author}{\bibfnamefont{M.}~\bibnamefont{Prakash}},
  \bibinfo{journal}{Astrophys. J.} \textbf{\bibinfo{volume}{550}},
  \bibinfo{pages}{426} (\bibinfo{year}{2001}), \eprint{astro-ph/0002232}.

\bibitem[{\citenamefont{Lattimer and Prakash}(2004)}]{Lattimer:2004pg}
\bibinfo{author}{\bibfnamefont{J.~M.} \bibnamefont{Lattimer}} \bibnamefont{and}
  \bibinfo{author}{\bibfnamefont{M.}~\bibnamefont{Prakash}},
  \bibinfo{journal}{Science} \textbf{\bibinfo{volume}{304}},
  \bibinfo{pages}{536} (\bibinfo{year}{2004}), \eprint{astro-ph/0405262}.

\bibitem[{\citenamefont{Lattimer and Prakash}(2007)}]{Lattimer:2006xb}
\bibinfo{author}{\bibfnamefont{J.~M.} \bibnamefont{Lattimer}} \bibnamefont{and}
  \bibinfo{author}{\bibfnamefont{M.}~\bibnamefont{Prakash}},
  \bibinfo{journal}{Phys. Rept.} \textbf{\bibinfo{volume}{442}},
  \bibinfo{pages}{109} (\bibinfo{year}{2007}), \eprint{astro-ph/0612440}.

\bibitem[{\citenamefont{{Lindblom}}(1992)}]{Lindblom:1992}
\bibinfo{author}{\bibfnamefont{L.}~\bibnamefont{{Lindblom}}},
  \bibinfo{journal}{\apj} \textbf{\bibinfo{volume}{398}}, \bibinfo{pages}{569}
  (\bibinfo{year}{1992}).

\bibitem[{\citenamefont{Mueller and Serot}(1996)}]{Mueller:1996pm}
\bibinfo{author}{\bibfnamefont{H.}~\bibnamefont{Mueller}} \bibnamefont{and}
  \bibinfo{author}{\bibfnamefont{B.~D.} \bibnamefont{Serot}},
  \bibinfo{journal}{Nucl. Phys.} \textbf{\bibinfo{volume}{A606}},
  \bibinfo{pages}{508} (\bibinfo{year}{1996}), \eprint{nucl-th/9603037}.

\bibitem[{\citenamefont{Lalazissis et~al.}(1997)\citenamefont{Lalazissis,
  Konig, and Ring}}]{Lalazissis:1996rd}
\bibinfo{author}{\bibfnamefont{G.~A.} \bibnamefont{Lalazissis}},
  \bibinfo{author}{\bibfnamefont{J.}~\bibnamefont{Konig}}, \bibnamefont{and}
  \bibinfo{author}{\bibfnamefont{P.}~\bibnamefont{Ring}},
  \bibinfo{journal}{Phys. Rev.} \textbf{\bibinfo{volume}{C55}},
  \bibinfo{pages}{540} (\bibinfo{year}{1997}), \eprint{nucl-th/9607039}.

\bibitem[{\citenamefont{Lalazissis et~al.}(1999)\citenamefont{Lalazissis,
  Raman, and Ring}}]{Lalazissis:1999}
\bibinfo{author}{\bibfnamefont{G.~A.} \bibnamefont{Lalazissis}},
  \bibinfo{author}{\bibfnamefont{S.}~\bibnamefont{Raman}}, \bibnamefont{and}
  \bibinfo{author}{\bibfnamefont{P.}~\bibnamefont{Ring}}, \bibinfo{journal}{At.
  Data Nucl. Data Tables} \textbf{\bibinfo{volume}{71}}, \bibinfo{pages}{1}
  (\bibinfo{year}{1999}).

\bibitem[{\citenamefont{Todd-Rutel and Piekarewicz}(2005)}]{Todd-Rutel:2005fa}
\bibinfo{author}{\bibfnamefont{B.~G.} \bibnamefont{Todd-Rutel}}
  \bibnamefont{and}
  \bibinfo{author}{\bibfnamefont{J.}~\bibnamefont{Piekarewicz}},
  \bibinfo{journal}{Phys. Rev. Lett} \textbf{\bibinfo{volume}{95}},
  \bibinfo{pages}{122501} (\bibinfo{year}{2005}), \eprint{nucl-th/0504034}.

\bibitem[{\citenamefont{Piekarewicz}(2009)}]{Piekarewicz:2009gb}
\bibinfo{author}{\bibfnamefont{J.}~\bibnamefont{Piekarewicz}}
  (\bibinfo{year}{2009}), \eprint{0912.5103}.

\bibitem[{\citenamefont{Baym et~al.}(1971)\citenamefont{Baym, Pethick, and
  Sutherland}}]{Baym:1971pw}
\bibinfo{author}{\bibfnamefont{G.}~\bibnamefont{Baym}},
  \bibinfo{author}{\bibfnamefont{C.}~\bibnamefont{Pethick}}, \bibnamefont{and}
  \bibinfo{author}{\bibfnamefont{P.}~\bibnamefont{Sutherland}},
  \bibinfo{journal}{Astrophys. J.} \textbf{\bibinfo{volume}{170}},
  \bibinfo{pages}{299} (\bibinfo{year}{1971}).

\bibitem[{\citenamefont{Ravenhall et~al.}(1983)\citenamefont{Ravenhall,
  Pethick, and Wilson}}]{Ravenhall:1983uh}
\bibinfo{author}{\bibfnamefont{D.~G.} \bibnamefont{Ravenhall}},
  \bibinfo{author}{\bibfnamefont{C.~J.} \bibnamefont{Pethick}},
  \bibnamefont{and} \bibinfo{author}{\bibfnamefont{J.~R.}
  \bibnamefont{Wilson}}, \bibinfo{journal}{Phys. Rev. Lett.}
  \textbf{\bibinfo{volume}{50}}, \bibinfo{pages}{2066} (\bibinfo{year}{1983}).

\bibitem[{\citenamefont{Hashimoto et~al.}(1984)\citenamefont{Hashimoto, Seki,
  and Yamada}}]{Hashimoto:1984}
\bibinfo{author}{\bibfnamefont{M.}~\bibnamefont{Hashimoto}},
  \bibinfo{author}{\bibfnamefont{H.}~\bibnamefont{Seki}}, \bibnamefont{and}
  \bibinfo{author}{\bibfnamefont{M.}~\bibnamefont{Yamada}},
  \bibinfo{journal}{Prog. Theor. Phys.} \textbf{\bibinfo{volume}{71}},
  \bibinfo{pages}{320} (\bibinfo{year}{1984}).

\bibitem[{\citenamefont{Lorenz et~al.}(1993)\citenamefont{Lorenz, Ravenhall,
  and Pethick}}]{Lorenz:1992zz}
\bibinfo{author}{\bibfnamefont{C.~P.} \bibnamefont{Lorenz}},
  \bibinfo{author}{\bibfnamefont{D.~G.} \bibnamefont{Ravenhall}},
  \bibnamefont{and} \bibinfo{author}{\bibfnamefont{C.~J.}
  \bibnamefont{Pethick}}, \bibinfo{journal}{Phys. Rev. Lett.}
  \textbf{\bibinfo{volume}{70}}, \bibinfo{pages}{379} (\bibinfo{year}{1993}).

\bibitem[{\citenamefont{Horowitz
  et~al.}(2004{\natexlab{a}})\citenamefont{Horowitz, Perez-Garcia, and
  Piekarewicz}}]{Horowitz:2004yf}
\bibinfo{author}{\bibfnamefont{C.~J.} \bibnamefont{Horowitz}},
  \bibinfo{author}{\bibfnamefont{M.~A.} \bibnamefont{Perez-Garcia}},
  \bibnamefont{and}
  \bibinfo{author}{\bibfnamefont{J.}~\bibnamefont{Piekarewicz}},
  \bibinfo{journal}{Phys. Rev.} \textbf{\bibinfo{volume}{C69}},
  \bibinfo{pages}{045804} (\bibinfo{year}{2004}{\natexlab{a}}),
  \eprint{astro-ph/0401079}.

\bibitem[{\citenamefont{Horowitz
  et~al.}(2004{\natexlab{b}})\citenamefont{Horowitz, Perez-Garcia, Carriere,
  Berry, and Piekarewicz}}]{Horowitz:2004pv}
\bibinfo{author}{\bibfnamefont{C.~J.} \bibnamefont{Horowitz}},
  \bibinfo{author}{\bibfnamefont{M.~A.} \bibnamefont{Perez-Garcia}},
  \bibinfo{author}{\bibfnamefont{J.}~\bibnamefont{Carriere}},
  \bibinfo{author}{\bibfnamefont{D.~K.} \bibnamefont{Berry}}, \bibnamefont{and}
  \bibinfo{author}{\bibfnamefont{J.}~\bibnamefont{Piekarewicz}},
  \bibinfo{journal}{Phys. Rev.} \textbf{\bibinfo{volume}{C70}},
  \bibinfo{pages}{065806} (\bibinfo{year}{2004}{\natexlab{b}}),
  \eprint{astro-ph/0409296}.

\bibitem[{\citenamefont{Horowitz et~al.}(2005)\citenamefont{Horowitz,
  Perez-Garcia, Berry, and Piekarewicz}}]{Horowitz:2005zb}
\bibinfo{author}{\bibfnamefont{C.~J.} \bibnamefont{Horowitz}},
  \bibinfo{author}{\bibfnamefont{M.~A.} \bibnamefont{Perez-Garcia}},
  \bibinfo{author}{\bibfnamefont{D.~K.} \bibnamefont{Berry}}, \bibnamefont{and}
  \bibinfo{author}{\bibfnamefont{J.}~\bibnamefont{Piekarewicz}},
  \bibinfo{journal}{Phys. Rev.} \textbf{\bibinfo{volume}{C72}},
  \bibinfo{pages}{035801} (\bibinfo{year}{2005}), \eprint{nucl-th/0508044}.

\bibitem[{\citenamefont{Link et~al.}(1999)\citenamefont{Link, Epstein, and
  Lattimer}}]{Link:1999ca}
\bibinfo{author}{\bibfnamefont{B.}~\bibnamefont{Link}},
  \bibinfo{author}{\bibfnamefont{R.~I.} \bibnamefont{Epstein}},
  \bibnamefont{and} \bibinfo{author}{\bibfnamefont{J.~M.}
  \bibnamefont{Lattimer}}, \bibinfo{journal}{Phys. Rev. Lett.}
  \textbf{\bibinfo{volume}{83}}, \bibinfo{pages}{3362} (\bibinfo{year}{1999}),
  \eprint{astro-ph/9909146}.

\bibitem[{\citenamefont{Carriere et~al.}(2003)\citenamefont{Carriere, Horowitz,
  and Piekarewicz}}]{Carriere:2002bx}
\bibinfo{author}{\bibfnamefont{J.}~\bibnamefont{Carriere}},
  \bibinfo{author}{\bibfnamefont{C.~J.} \bibnamefont{Horowitz}},
  \bibnamefont{and}
  \bibinfo{author}{\bibfnamefont{J.}~\bibnamefont{Piekarewicz}},
  \bibinfo{journal}{Astrophys. J.} \textbf{\bibinfo{volume}{593}},
  \bibinfo{pages}{463} (\bibinfo{year}{2003}), \eprint{nucl-th/0211015}.

\bibitem[{\citenamefont{Serot and Walecka}(1986)}]{Serot:1984ey}
\bibinfo{author}{\bibfnamefont{B.~D.} \bibnamefont{Serot}} \bibnamefont{and}
  \bibinfo{author}{\bibfnamefont{J.~D.} \bibnamefont{Walecka}},
  \bibinfo{journal}{Adv. Nucl. Phys.} \textbf{\bibinfo{volume}{16}},
  \bibinfo{pages}{1} (\bibinfo{year}{1986}).

\bibitem[{\citenamefont{Serot and Walecka}(1997)}]{Serot:1997xg}
\bibinfo{author}{\bibfnamefont{B.~D.} \bibnamefont{Serot}} \bibnamefont{and}
  \bibinfo{author}{\bibfnamefont{J.~D.} \bibnamefont{Walecka}},
  \bibinfo{journal}{Int. J. Mod. Phys.} \textbf{\bibinfo{volume}{E6}},
  \bibinfo{pages}{515} (\bibinfo{year}{1997}), \eprint{nucl-th/9701058}.

\bibitem[{\citenamefont{Horowitz and
  Piekarewicz}(2001{\natexlab{a}})}]{Horowitz:2000xj}
\bibinfo{author}{\bibfnamefont{C.~J.} \bibnamefont{Horowitz}} \bibnamefont{and}
  \bibinfo{author}{\bibfnamefont{J.}~\bibnamefont{Piekarewicz}},
  \bibinfo{journal}{Phys. Rev. Lett.} \textbf{\bibinfo{volume}{86}},
  \bibinfo{pages}{5647} (\bibinfo{year}{2001}{\natexlab{a}}),
  \eprint{astro-ph/0010227}.

\bibitem[{\citenamefont{Todd and Piekarewicz}(2003)}]{Todd:2003xs}
\bibinfo{author}{\bibfnamefont{B.~G.} \bibnamefont{Todd}} \bibnamefont{and}
  \bibinfo{author}{\bibfnamefont{J.}~\bibnamefont{Piekarewicz}},
  \bibinfo{journal}{Phys. Rev.} \textbf{\bibinfo{volume}{C67}},
  \bibinfo{pages}{044317} (\bibinfo{year}{2003}), \eprint{nucl-th/0301092}.

\bibitem[{\citenamefont{Boguta and Bodmer}(1977)}]{Boguta:1977xi}
\bibinfo{author}{\bibfnamefont{J.}~\bibnamefont{Boguta}} \bibnamefont{and}
  \bibinfo{author}{\bibfnamefont{A.~R.} \bibnamefont{Bodmer}},
  \bibinfo{journal}{Nucl. Phys.} \textbf{\bibinfo{volume}{A292}},
  \bibinfo{pages}{413} (\bibinfo{year}{1977}).

\bibitem[{\citenamefont{Youngblood et~al.}(1999)\citenamefont{Youngblood,
  Clark, and Lui}}]{Youngblood:1999}
\bibinfo{author}{\bibfnamefont{D.~H.} \bibnamefont{Youngblood}},
  \bibinfo{author}{\bibfnamefont{H.~L.} \bibnamefont{Clark}}, \bibnamefont{and}
  \bibinfo{author}{\bibfnamefont{Y.-W.} \bibnamefont{Lui}},
  \bibinfo{journal}{Phys. Rev. Lett.} \textbf{\bibinfo{volume}{82}},
  \bibinfo{pages}{691} (\bibinfo{year}{1999}).

\bibitem[{\citenamefont{Horowitz and
  Piekarewicz}(2001{\natexlab{b}})}]{Horowitz:2001ya}
\bibinfo{author}{\bibfnamefont{C.~J.} \bibnamefont{Horowitz}} \bibnamefont{and}
  \bibinfo{author}{\bibfnamefont{J.}~\bibnamefont{Piekarewicz}},
  \bibinfo{journal}{Phys. Rev.} \textbf{\bibinfo{volume}{C64}},
  \bibinfo{pages}{062802} (\bibinfo{year}{2001}{\natexlab{b}}),
  \eprint{nucl-th/0108036}.

\bibitem[{\citenamefont{Horowitz et~al.}(2001)\citenamefont{Horowitz, Pollock,
  Souder, and Michaels}}]{Horowitz:1999fk}
\bibinfo{author}{\bibfnamefont{C.~J.} \bibnamefont{Horowitz}},
  \bibinfo{author}{\bibfnamefont{S.~J.} \bibnamefont{Pollock}},
  \bibinfo{author}{\bibfnamefont{P.~A.} \bibnamefont{Souder}},
  \bibnamefont{and} \bibinfo{author}{\bibfnamefont{R.}~\bibnamefont{Michaels}},
  \bibinfo{journal}{Phys. Rev.} \textbf{\bibinfo{volume}{C63}},
  \bibinfo{pages}{025501} (\bibinfo{year}{2001}), \eprint{nucl-th/9912038}.

\bibitem[{\citenamefont{Kumar et~al.}(2005)\citenamefont{Kumar, Michaels,
  Souder, and Urciuoli}}]{Michaels:2005}
\bibinfo{author}{\bibfnamefont{K.}~\bibnamefont{Kumar}},
  \bibinfo{author}{\bibfnamefont{R.}~\bibnamefont{Michaels}},
  \bibinfo{author}{\bibfnamefont{P.~A.} \bibnamefont{Souder}},
  \bibnamefont{and} \bibinfo{author}{\bibfnamefont{G.~M.}
  \bibnamefont{Urciuoli}} (\bibinfo{year}{2005}),
  \urlprefix\url{http://hallaweb.jlab.org/parity/prex}.

\bibitem[{\citenamefont{Guver et~al.}(2008)\citenamefont{Guver, Ozel,
  Cabrera-Lavers, and Wroblewski}}]{Guver:2008gc}
\bibinfo{author}{\bibfnamefont{T.}~\bibnamefont{Guver}},
  \bibinfo{author}{\bibfnamefont{F.}~\bibnamefont{Ozel}},
  \bibinfo{author}{\bibfnamefont{A.}~\bibnamefont{Cabrera-Lavers}},
  \bibnamefont{and}
  \bibinfo{author}{\bibfnamefont{P.}~\bibnamefont{Wroblewski}}
  (\bibinfo{year}{2008}), \eprint{0811.3979}.

\bibitem[{\citenamefont{Ozel et~al.}(2009)\citenamefont{Ozel, Guver, and
  Psaltis}}]{Ozel:2008kb}
\bibinfo{author}{\bibfnamefont{F.}~\bibnamefont{Ozel}},
  \bibinfo{author}{\bibfnamefont{T.}~\bibnamefont{Guver}}, \bibnamefont{and}
  \bibinfo{author}{\bibfnamefont{D.}~\bibnamefont{Psaltis}},
  \bibinfo{journal}{Astrophys. J.} \textbf{\bibinfo{volume}{693}},
  \bibinfo{pages}{1775} (\bibinfo{year}{2009}), \eprint{0810.1521}.

\bibitem[{\citenamefont{Guver et~al.}(2010)\citenamefont{Guver, Wroblewski,
  Camarota, and Ozel}}]{Guver:2010td}
\bibinfo{author}{\bibfnamefont{T.}~\bibnamefont{Guver}},
  \bibinfo{author}{\bibfnamefont{P.}~\bibnamefont{Wroblewski}},
  \bibinfo{author}{\bibfnamefont{L.}~\bibnamefont{Camarota}}, \bibnamefont{and}
  \bibinfo{author}{\bibfnamefont{F.}~\bibnamefont{Ozel}}
  (\bibinfo{year}{2010}), \eprint{1002.3825}.

\bibitem[{\citenamefont{Schwenk and Pethick}(2005)}]{Schwenk:2005ka}
\bibinfo{author}{\bibfnamefont{A.}~\bibnamefont{Schwenk}} \bibnamefont{and}
  \bibinfo{author}{\bibfnamefont{C.~J.} \bibnamefont{Pethick}},
  \bibinfo{journal}{Phys. Rev. Lett.} \textbf{\bibinfo{volume}{95}},
  \bibinfo{pages}{160401} (\bibinfo{year}{2005}), \eprint{nucl-th/0506042}.

\bibitem[{\citenamefont{Gezerlis and Carlson}(2009)}]{Gezerlis:2009iw}
\bibinfo{author}{\bibfnamefont{A.}~\bibnamefont{Gezerlis}} \bibnamefont{and}
  \bibinfo{author}{\bibfnamefont{J.}~\bibnamefont{Carlson}}
  (\bibinfo{year}{2009}), \eprint{0911.3907}.

\bibitem[{\citenamefont{Piekarewicz}(2007)}]{Piekarewicz:2007dx}
\bibinfo{author}{\bibfnamefont{J.}~\bibnamefont{Piekarewicz}},
  \bibinfo{journal}{Phys. Rev.} \textbf{\bibinfo{volume}{C76}},
  \bibinfo{pages}{064310} (\bibinfo{year}{2007}), \eprint{0709.2699}.

\end{thebibliography}
\end{document}